\documentclass[ngerman,final,3p,times,twocolumn]{elsarticle}

\usepackage{graphicx}
\usepackage{amssymb}

\usepackage[utf8]{inputenc}

\usepackage{xfrac}

\usepackage{hyperref}
\hypersetup{
    colorlinks,
    citecolor=blue,
    filecolor=blue,
    linkcolor=blue,
    urlcolor=blue
}

\usepackage{xcolor}

\usepackage{lineno}

\begin{document}
\newcommand{\yl}[1]{\textcolor{red}{#1}}

\begin{frontmatter}


\title{Storage, Accumulation and Deceleration of Secondary Beams\\ for Nuclear Astrophysics}

\author[gsi]{\mbox{J. Glorius}}
\ead{j.glorius@gsi.de}
\author[gsi]{\mbox{Yu. A. Litvinov}}
\author[edin]{\mbox{M. Aliotta}}
\author[gsi]{\mbox{F. Amjad}}
\author[guf]{\mbox{B. Br\"uckner}}
\author[edin]{\mbox{C.G. Bruno}}
\author[gsi,imp]{\mbox{R. Chen}}
\author[edin]{\mbox{T. Davinson}}
\author[guf,gsi]{\mbox{S.F. Dellmann}}
\author[gsi,jlu]{\mbox{T. Dickel}}
\author[triumf,uvic]{\mbox{I. Dillmann}}
\author[guf]{\mbox{P. Erbacher}}
\author[gsi]{\mbox{O. Forstner}}
\author[gsi]{\mbox{H. Geissel}}
\author[triumf]{\mbox{C.J. Griffin}}
\author[gsi]{\mbox{R. Grisenti}}
\author[gsi]{\mbox{A. Gumberidze}}
\author[gsi]{\mbox{E. Haettner}}
\author[gsi]{\mbox{R. Hess}}
\author[gsi,jlu]{\mbox{P.-M. Hillenbrand}}
\author[gsi]{\mbox{C. Hornung}}
\author[gsi]{\mbox{R. Joseph}}
\author[lp2i]{\mbox{B. Jurado}}
\author[gsi]{\mbox{E. Kazanseva}}
\author[gsi]{\mbox{R. Kn\"obel}}
\author[gsi]{\mbox{D. Kostyleva}}
\author[gsi]{\mbox{C. Kozhuharov}}
\author[gsi]{\mbox{N. Kuzminchuk}}
\author[aac]{\mbox{C. Langer}}
\author[triumf]{\mbox{G. Leckenby}}
\author[edin]{\mbox{C. Lederer-Woods}}
\author[gsi]{\mbox{M. Lestinsky}}
\author[gsi]{\mbox{S. Litvinov}}
\author[gsi]{\mbox{B. L\"oher}}
\author[gsi]{\mbox{B. Lorenz}}
\author[guf]{\mbox{E. Lorenz}}
\author[edin]{\mbox{J. Marsh}}
\author[gsi]{\mbox{E. Menz}}
\author[gsi]{\mbox{T. Morgenroth}}
\author[gsi]{\mbox{I. Mukha}}
\author[gsi]{\mbox{N. Petridis}}
\author[gsi]{\mbox{U. Popp}}
\author[tud]{\mbox{A. Psaltis}}
\author[gsi]{\mbox{S. Purushothaman}}
\author[guf]{\mbox{R. Reifarth}}
\author[gsi]{\mbox{E. Rocco}}
\author[gsi]{\mbox{P. Roy}}
\author[gsi,aac]{\mbox{M. S. Sanjari}}
\author[gsi,jlu]{\mbox{C. Scheidenberger}}
\author[lp2i]{\mbox{M. Sguazzin}}
\author[gsi,edin]{\mbox{R.S. Sidhu}}
\author[gsi]{\mbox{U. Spillmann}}
\author[gsi]{\mbox{M. Steck}}
\author[gsi,hij]{\mbox{T. St\"ohlker}}
\author[lp2i]{\mbox{J.A. Swartz}}
\author[gsi,riken]{\mbox{Y.K. Tanaka}}
\author[gsi,tud]{\mbox{H. T\"ornqvist}}
\author[gsi]{\mbox{L. Varga}}
\author[guf]{\mbox{D. Vescovi}}
\author[gsi]{\mbox{H. Weick}}
\author[guf]{\mbox{M. Weigand}}
\author[edin]{\mbox{P. J. Woods}}
\author[usai,utsu]{\mbox{T. Yamaguchi}}
\author[gsi]{\mbox{J. Zhao}}

\address[gsi]{GSI Helmholtzzentrum f\"ur Schwerionenforschung GmbH, Darmstadt, Germany}
\address[edin]{{University of Edinburgh, Edinburgh, United Kingdom}}
\address[guf]{{Goethe Universit\"at, Frankfurt am Main, Germany}}
\address[imp]{{Institute of Modern Physics, Chinese Academy of Sciences, Lanzhou, China}}
\address[triumf]{{TRIUMF, Vancouver, British Columbia, Canada}}
\address[uvic]{{Department of Physics and Astronomy, University of Victoria, Victoria, Canada}}
\address[jlu]{{Justus-Liebig Universit\"at, Gie{\ss}en, Germany}}
\address[lp2i]{{LP2I Bordeaux, CNRS-IN2P3, Gradignan, France}}
\address[aac]{{University of Applied Sciences, Aachen, Germany}}
\address[hij]{{Helmholtz-Institut Jena, Jena, Germany}}
\address[tud]{{Technische Universit\"at Darmstadt, Darmstadt, Germany}}
\address[riken]{{High Energy Nuclear Physics Laboratory, RIKEN, Saitama, Japan}}
\address[usai]{{Department of Physics, Saitama University, Saitama, Japan}}
\address[utsu]{{Tomonaga Center for the History of the Universe, University of Tsukuba, Ibaraki, Japan}}

\begin{abstract}
Low-energy investigations on rare ion beams are often limited by the available intensity and purity of the ion species in focus. Here, we present the first application of a technique that combines in-flight production at relativistic energies with subsequent secondary beam storage, accumulation and finally deceleration to the energy of interest. 
Using the FRS and ESR facilities at GSI, this scheme was pioneered to provide a secondary beam of $^{118}$Te$^{52+}$ for the measurement of nuclear proton-capture at energies of 6 and 7 MeV/u. The technique provided stored beam intensities of about $10^6$ ions at high purity and brilliance, representing a major step towards low-energy nuclear physics studies using rare ion beams.
\end{abstract}


\end{frontmatter}


\section{Introduction}
\label{sec:intro}

Inverse kinematic experiments with stored and cooled secondary beams have long been identified as a very promising approach to explore properties and reactions beyond nuclear stability, {\em e.g.} to address the large nuclear uncertainties in astrophysical nucleosynthesis \cite{henning97, bertulani97}. However, the required advances in detector technology, accelerator techniques and related fields only became available in the last decade. The emergence of ultra-high vacuum detection systems, efficient post-deceleration of secondary beams and powerful stored-beam diagnostics eventually triggered a new era of experimental endeavors addressing direct reaction studies at low-energies \cite{Grieser-2012s, Lestinsky-2016s, Glorius-2023a}.

The Experimental Storage Ring (ESR) at GSI \cite{Franzke-1987, Geissel-1992s, Litvinov-2013s} plays a pioneering role for this kind of experimental scheme. The flexible heavy ion storage ring is able to store any ion beam from protons to uranium, including secondary beams produced in-flight in the Fragment Separator FRS \cite{Geissel-1992ds, Franzke-2008}. Possible beam energies of stored heavy ions range from 550 MeV/u down to about 4 MeV/u, while the lower end of this range is usually reached by means of deceleration. Further indispensable features of the ring are beam cooling, a large momentum acceptance of $\pm 1.5\%$, an internal gas target, non-destructive beam diagnostics as well as a large variety of beam manipulations such as beam bunching, global and local orbit shifts or precise acceptance trimming by mechanical scrapers. See \cite{Franzke-1987, Steck-2020} for further details.

With this combination of radioactive-ion production, separation, storage and versatile beam handling features, the facility is the perfect host to pioneer and establish new experimental schemes for low-energy nuclear structure and nuclear astrophysics. Both fields nowadays commonly deal with small production cross sections of the nuclei of interest and high requirements on achievable luminosity and resolution. These boundary conditions also apply to the recently established experimental approach to measure proton-induced reaction cross sections in the ring environment. While the final goal of this technique is to provide cross section data in the realm of unstable nuclei, {\em e.g.} for modelling explosive nucleosynthesis, it has initially been developed with stable heavy ion beams \cite{Zhong-2010s, Mei-2015s,Glorius-2019s}. Only lately the technique reached a state of maturity that enabled the first application to exotic, secondary beams \cite{Dellmann-2022s, Varga2021phd}. 

In this paper we take the opportunity to report on the performance of low-energy storage of secondary beams in the ESR in a joint operation with the FRS. The focus will be on a fragment beam of $^{118}_{52}$Te$^{52+}$ ($t_{\text{\sfrac{1}{2}}} = 6.00(2)~\text{d}$\,\cite{NNDC}) produced for the first measurement of a proton-capture cross section on a stored radionuclide. See \cite{Dellmann-2022s} for details on the experiment. 

\section{In-flight Production of $^{118}$Te}

In the FRS the production of neutron-deficient secondary beams, such as $^{118}$Te, is usually accomplished by in-flight projectile fragmentation on a production target of beryllium. 
The fragment cocktail beam that is generated in the process is subsequently filtered in two magnetic separation stages assisted by an intermediate energy-loss in a special degrader. 
Further details of the concept can be found elsewhere \cite{Franzke-2008, Geissel-1995d}. 
The beam setup optimized for injection of $^{118}$Te$^{52+}$ into the ESR started with a primary beam of $^{124}$Xe at 550\,MeV/u incident on a $^9$Be target with a thickness of about 2.5\,g/cm$^2$. 
After passage through the production target, a niobium stripper foil of 220\,mg/cm$^2$ and a plastic degrader sheet (2\,mm), the secondary beam reached an energy of 400\,MeV/u for injection into ESR. The two additional stripping stages ensured a fully-ionized configuration of the ions, while other charge states were strongly suppressed.

As a point of reference, the effective fragment yield for bare ions of $^{118}$Te stored in the ESR can be quantified as $3.5\cdot10^5$ pps with a primary intensity of about $3.5\cdot10^9$\,pps for $^{124}$Xe from the heavy-ion synchrotron SIS18\,\cite{Blasche-1985}. This corresponds to a production efficiency of $10^{-4}$ which is compatible within a factor of two with predictions based on the LISE++ transport code\,\cite{tarasov2008_LISE}. 
The injection into the ESR is the main unknown in the simulations, which is parametrized by cuts in transverse position and momentum. 
In the present context, the default injection parameters implemented in the version 16.1 of the code have been used.
The issue of contaminants, i.e. hydrogen-like fragments close to $^{118}$Te$^{52+}$ in $m/q$, will be discussed in the following section.

\section{Storage and Manipulation of the $^{118}$Te$^{52+}$ Fragment Beams}

In order to increase the number of $^{118}$Te$^{52+}$ ions stored in the ring, about 20 consecutive injections were accumulated. This was accomplished using a beam stacking technique established about 10 years ago with a focus on radioactive beams \cite{Nolden-2013s}. 
The full concept to store pure, rare ion beams at energies below 10 MeV/u achieved here, consists of the following steps:

\begin{enumerate}
    \item injection of the hot fragment beam from the FRS onto an outer ring orbit at 400\,MeV/u;
    \item stochastic pre-cooling and bunching of the beam on the injection orbit;
    \item selective displacement of $^{118}$Te$^{52+}$ to an inner orbit by deceleration with a rapid frequency (RF) at fixed magnetic field;
    \item accumulation of $^{118}$Te$^{52+}$ ions on an inner stacking orbit by means of electron cooling;
    \item repetition of steps 1 to 4 until desired intensity is reached;   
    \item deceleration of bunched, accumulated beam to low energies;
    \item continuous electron cooling of the decelerated, coasting beam at the low energy of interest.
\end{enumerate}

\begin{figure}[t!]
    \centering
    \includegraphics[width=\columnwidth]{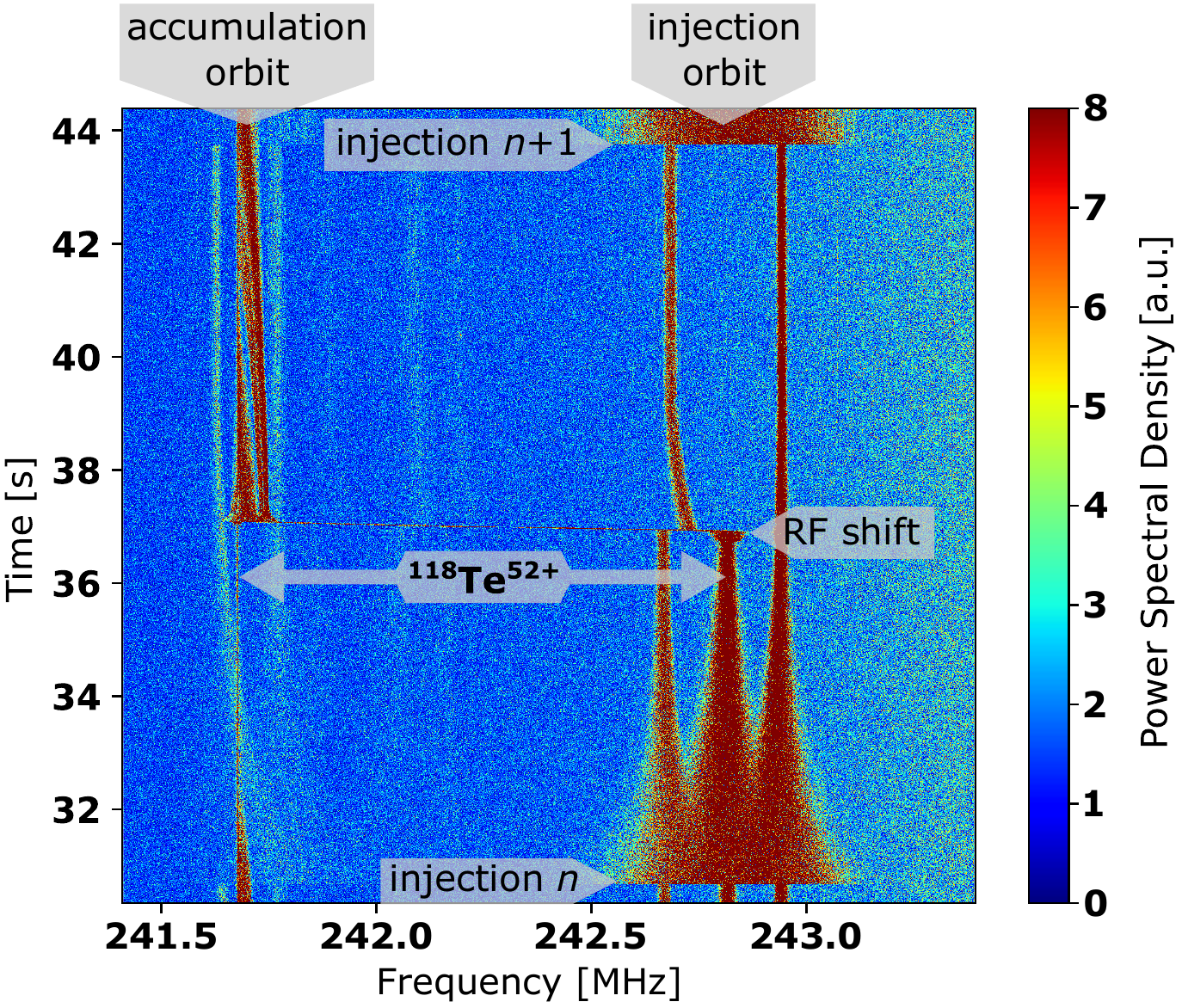}
    \caption{Single injection and accumulation procedure in the frequency domain recorded by time-resolved Schottky spectrometry. Displayed is the 125$^{\rm{ th}}$ harmonic of the beam revolution frequency. For details see text. Note the logarithmic intensity scale.}
    \label{fig:accu}
\end{figure}

In figure \ref{fig:accu} a single injection of the accumulation procedure is visualized in the frequency domain as measured by time-resolved Schottky spectroscopy\,\cite{Litvinov-2004as, Nolden-2011s, Kienle-2013s, Ozturk-2019s}. 
The circumference of the ESR is 108\,m and thus the revolution frequency of the stored ions at 400\,MeV/u is about 2\,MHz.
Displayed is the 125$^{\rm{th}}$ harmonic of the revolution frequency. For details on acquisition, signal processing and analysis see\,\cite{Nolden-2011s, Sanjari-2020s, sanjari-2023}.
The color code represents the logarithm of the stored beam intensity.

Ion beams produced in-flight suffer a large momentum spread $\Delta p/p$, mainly caused by the nuclear reaction kinematics at the production\,\cite{Geissel-1995d}. Therefore, a special concept for cooling of the secondary beam in the ESR was decisive.

For the initial stochastic cooling the Palmer technique was applied, see {\em e.g.} \cite{nolden2004}, which provided fast pre-cooling of the hot fragments to about $\Delta p/p\sim10^{-4}$ within 3-4 seconds. 
The stochastic cooling operates at a fixed beam energy of 400\,MeV/u and
acts in a limited frequency bandwidth, corresponding to about 242.5 to 243.0\,MHz in figure \ref{fig:accu}. 
It forces all stored ions to the same orbit independently on their mass-over-charge ratio, $m/q$.
It is optimized for $^{118}$Te$^{52+}$ which is cooled at about 242.8\,MHz. 
According to simulations, contaminant ions, such as  $^{123}$Xe$^{54+}$, $^{120}$I$^{53+}$ or $^{116}$Sb$^{51+}$, are transmitted and can be injected into the ESR.
The stochastic cooling acts also on these ions which, however, cannot be properly cooled as their $m/q$ values and the cooling orbit do not match. These ions are seen in figure \ref{fig:accu} at about 242.7 and 242.9\,MHz.

After the pre-cooling, the RF system of the ring has been utilized to pick-up the $^{118}$Te$^{52+}$ fragments and slightly reduce their energy.
The desired ions were shifted in a selective manner to an inner orbit at about 241.7\,MHz in figure \ref{fig:accu}.
The other species remained on the injection orbit and were kicked out with the subsequent injection.

The electron cooling is effective when the relative velocities of the cooler electrons and ions are small\,\cite{Steck-2020, Steck-2004} and in first order does not affect the ions stored on outer orbits at higher energy.
After the shift, the Te ions were placed very close to the accumulation orbit defined by the parameters of the electron cooler system, which was operating permanently. 
As a result, the displaced fragments were slowly brought into electron cooling equilibrium and merged with the already cooled $^{118}$Te$^{52+}$ beam on the accumulation orbit. 
On this orbit the beam is maintained and remains largely undisturbed by the injection kicker and/or the stochastic cooling system.

The faint lines visible in spectra in figure\,\ref{fig:accu} reveal that indeed some contaminants were also injected, stored and moved to the inner orbits. The electron cooling forces all ions to the same velocity and thus their revolution frequencies reflect their $m/q$ values\,\cite{Litvinov-2005bs}. Although the identification of these fragments shall in principle be possible, it is beyond the scope of this work. Figure\,\ref{fig:accu} shows that the overall scheme was carefully tailored to $^{118}$Te$^{52+}$, which as a result was the only ion species that got efficiently selected and accumulated. 

Directly after the accumulation the Te fragments were shifted to a central orbit by ramping down the ring magnetic field in preparation of the deceleration procedure to low energies. This removed the major part of the remaining contaminants still stored on outer orbits. 
The deceleration process implied capture of the beam of interest by an RF.
This process is similar to the shifting of $^{118}$Te$^{52+}$ beam during the accumulation and therefore adds further selectivity.
The beam energy was reduced by the RF accompanied by a simultaneous ramping down of the ESR magnetic fields and 
also the terminal high-voltage of the electron cooler. 
Slowing down from 400\,MeV/u to energies below about 10\,MeV/u requires to change the harmonic of the ring RF. 
Therefore, the deceleration was conducted in two steps with an intermediate electron cooling at 30\,MeV/u.

The deceleration itself is also prone to losses, especially for improperly cooled or non-centered beams. Starting with approximately $7\cdot10^6$ stored fragments after accumulation, about $1\cdot10^6$ cooled fragments were left at the final energy of 7 MeV/u targeted here. 
Since the operation of the stochastic cooling is fixed to the beam energy of 400\,MeV/u, only the electron cooling method is available at all other energies. If all other parameters are fixed, the achievable momentum spread increases with increasing beam intensity and with decreasing beam energy\,\cite{Poth-1990}. In our case $\Delta p/p \sim 10^{-5}$ can be assumed throughout the experiment\,\cite{Steck-1996, Steck-2004, Steck-2020}.

\begin{figure}[t!]
    \centering
    \includegraphics[width=0.95\columnwidth]{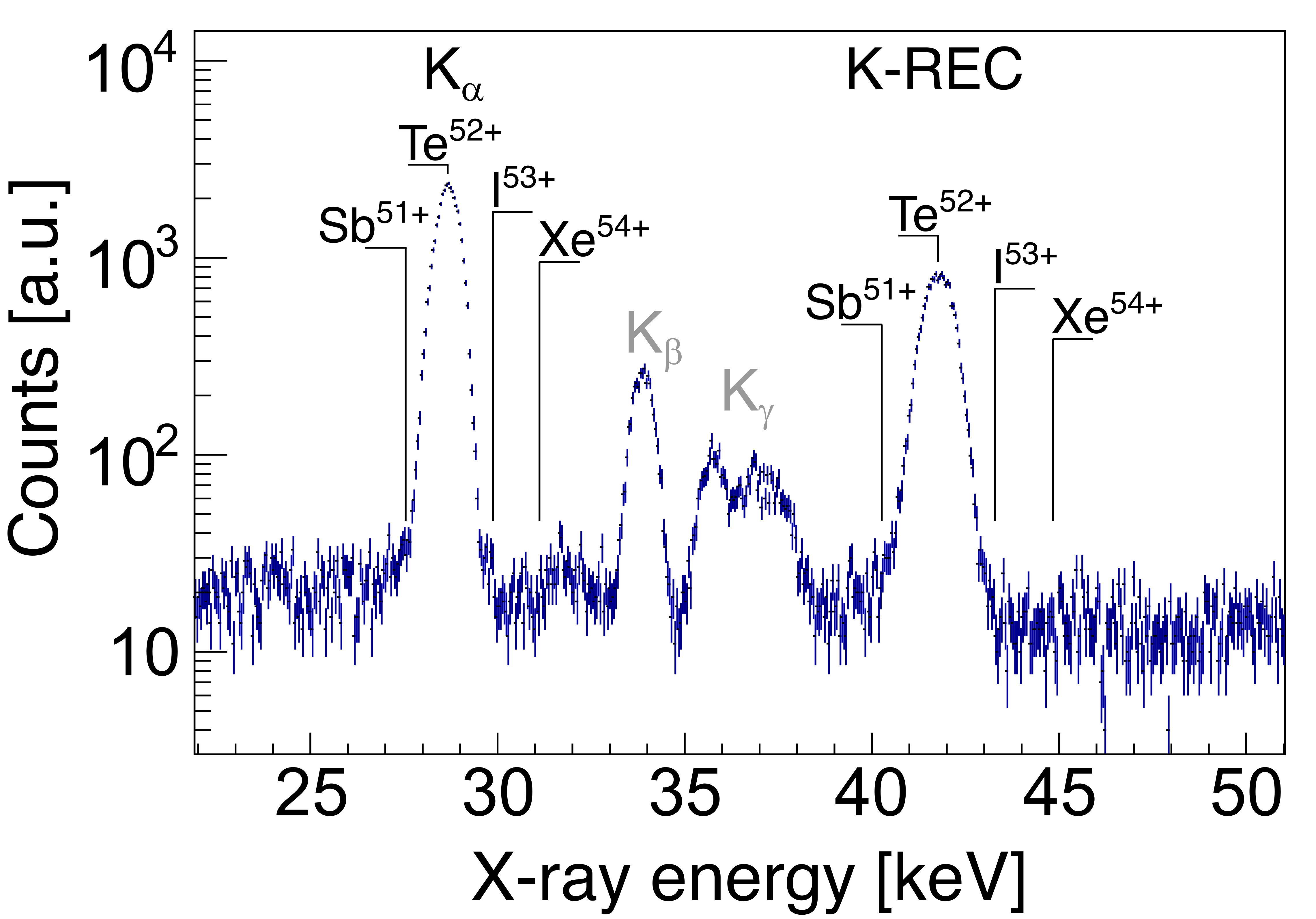}
    \caption{Energy spectrum of a germanium detector at the H$_2$ target under 90\textdegree, see text. All lines visible in this section can be clearly assigned to the characteristic X-ray or K-REC emission subsequent to electron capture on Te$^{52+}$ at the target. \label{fig:xray}}
\end{figure}

The purity of the low-energy beam was checked by means of X-ray spectroscopy at the internal target using hydrogen gas \cite{kuehnel2009}. The characteristic X-ray emission after electron capture at the target is a unique signature for each elemental species\,\cite{Gassner-2018s, Zhu-2022s} and is used here to infer the amount of residual beam impurities. We use binding energies extracted from the Dirac formalism to predict Doppler-corrected X-ray energies for $K_{\alpha}$ and also $K_{\text{REC}}$ emissions subsequent to electron capture on those fully-ionized species expected at high intensities from the FRS, see {\em e.g.} \cite{Eichler2007, Yerokhin-2015}. Since $K_{\alpha 1}$ and $K_{\alpha 2}$ lines are not resolved by the detector separately, we refer to $K_{\alpha 1}$ emission only. Higher order effects such as isotopic shifts or Lamb-shift are also neglected here. 

In figure \ref{fig:xray} a calibrated X-ray energy spectrum is shown as taken with a high-purity germanium detector placed at the target under 90{\textdegree} relative to the beam direction, including the calculated $K_{\alpha}$ and $K_{\text{REC}}$ energies for four relevant elements. All lines showing up in the spectrum match with the predictions for Te$^{52+}$ within about 50\,eV demonstrating a suppression of contaminants by at least two orders of magnitude.

The Te ions stored at 7 MeV/u exhibited a beam lifetime of about 1.5 seconds dominantly caused by atomic interactions with the residual gas, the H$_2$ target and the cooler electrons. A single beam cycle including accumulation, deceleration and measurement lasted about 360 seconds. 
With a measurement duration of 15 seconds, this implies a duty cycle of just over 4\%. Additionally, a setting for 6 MeV/u final energy has been established, resulting in a slightly lower intensity and a slightly shorter beam lifetime, and consequently a less efficient duty cycle.

\section{Conclusion}

For the first time, accumulation and deceleration of a secondary beam produced in-flight has been accomplished in the ESR storage ring. Although the overall duty cycle is merely a few percent, a pure $^{118}$Te$^{52+}$ beam at energies as low as 6 MeV/u has been stored and peak luminosities in excess of $10^{25}$\,cm$^{-2}$s$^{-1}$ facilitated a direct measurement of the proton-capture cross section using a hydrogen jet target. 

The complex beam preparation scheme applied here illustrates the versatility of the manipulation techniques available at heavy-ion storage rings.
In the present context, this enabled a flexible compensation of low production yields by means of beam accumulation. 
Furthermore, the adopted procedures acted as an efficient filter to further purify the beam of interest from fragment admixtures via selective beam manipulations. 
In future experiments, this additional filtering will allow relaxed requirements on the FRS setup, which in turn may lead to a higher transmission of secondary beams to the ESR.

The decisive bottleneck, however, is the deceleration efficiency. 
Here, the unwanted beam losses, which dramatically increase with decreasing beam energy, dictate the need for the highest possible degree of vacuum and rest-gas composition, in order to avoid as much as possible many-electron atoms\,\cite{Shevelko-2010}.
Further, the ramping speed and stabilization of the electron cooler high-voltage at the final energy is essential, because it determines the start time of the measurement in each cycle. 
In general, any delays in the preparation scheme, in particular at lowermost energies, where the losses are maximal, should be avoided.

The major conclusion of this work is that, while there is still potential to improve this technique in the future, the principle feasibility of exotic beam storage at low energy has been successfully demonstrated and thereby triggers a new era of nuclear physics studies.

\section*{Acknowledgement}
This project has received funding from the European Research Council (ERC) under the European Union's Horizon 2020 research and innovation programme (grant agreement No 682841 ``ASTRUm'' and No 884715 ``NECTAR'').
This work is further supported by 
the European Union (ChETEC-INFRA, project no. 101008324), 
the Helmholtz Forschungs-Akademie Hessen for FAIR (HFHF), 
the Federal Ministry of Education and Research (BMBF) under Grant No. 05P15RFFAA and 05P15RGFAA
and the State of Hesse within the Research Cluster ELEMENTS (Project ID 500/10.006).

\end{document}